\documentclass[lettersize,journal]{IEEEtran}
\usepackage{amsmath,amsfonts}
\usepackage{algorithmic}
\usepackage{algorithm}
\usepackage{array}
\usepackage[caption=false,font=normalsize,labelfont=sf,textfont=sf]{subfig}
\usepackage{textcomp}
\usepackage{stfloats}
\usepackage{url}
\usepackage{verbatim}
\usepackage{graphicx}
\usepackage{cite}
\usepackage{xcolor}
\usepackage{booktabs} 
\usepackage{multirow}
\newcommand{\B}[1]{\if#1\relax\bm
{#1}\else\mathbf{#1}\fi} 

\newtheorem{assumption}{Assumption}
\newtheorem{remark}{Remark}

\title{

A bioreactor-based architecture for \textit{in vivo} model-based and sim-to-real learning control of microbial consortium composition

}
\author{
    Sara Maria Brancato$^{1}$ \IEEEmembership{Student Member, IEEE}, 
    Davide Salzano$^{1}$, \IEEEmembership{Member, IEEE}, 
    Davide Fiore$^{2}$,
    Francesco De Lellis$^{1}$, \IEEEmembership{Member, IEEE},
    Giovanni Russo$^{3}$, \IEEEmembership{Senior Member, IEEE}, 
    Mario di Bernardo$^{1,4,*}$, \IEEEmembership{Fellow, IEEE}%
    \thanks{This work has been supported by the European Union- Next Generation EU, under PRIN 2022 PNRR, for the Project “Control of smart microbial communities for wastewater treatment.”}
    \thanks{$^1$ Sara Maria Brancato, Davide Salzano, Francesco De Lellis and Mario di Bernardo are with Department of Electrical Engineering and Information Technology,
    University of Naples Federico II, Via Claudio, 21, 80125, Naples, Italy, (email: saramaria.brancato@unina.it, davide.salzano@unina.it, francesco.delellis@unina.it, mario.dibernardo@unina.it)}%
    \thanks{$^2$ Davide Fiore is with Department of Mathematics and Applications ``R. Caccioppoli", University of Naples Federico II, Via Cintia, Monte S. Angelo I-80126, Naples, Italy (email: dvd.fiore@unina.it).}%
    \thanks{$^3$ Giovanni Russo is with Department of Information Engineering, Electrical Engineering and Applied Mathematics, 
    University of Salerno, Invariante 12/B, Via Giovanni Paolo II, 132, 84084, Fisciano, Italy (email: giova.russo@unisa.it).}%
    \thanks{$^4$ Mario di Bernardo is with Scuola Superiore Meridionale, Via Mezzocannone 4, 80134, Naples, Italy (email: mario.dibernardo@unina.it).}%
    \thanks{$^*$ Corresponding author.}%
    
}
\date{April 2025}

\begin{document}

\maketitle

\begin{abstract}
    Microbial consortia offer significant biotechnological advantages over monocultures for bioproduction. However, industrial deployment is hampered by the lack of scalable architectures to ensure stable coexistence between populations. Existing strategies rely on genetic modifications, which impose metabolic load, or environmental changes, which can reduce production. We present a versatile control architecture to regulate density and composition of a two-strain consortium without genetic engineering or drastic environmental changes. Our bioreactor-based control architecture comprises a mixing chamber where both strains are co-cultured and a reservoir sustaining the slower-growing strain. For both chambers we develop model-based and sim-to-real learning controllers. The control architecture is then validated \textit{in vivo} on a two-strain \textit{Escherichia coli} consortium, achieving precise and robust regulation of consortium density and composition, including tracking of time-varying references and recovery from perturbations.
\end{abstract}

\section{Introduction}

Microbial consortia are pervasive in nature, inhabiting diverse environments from the human body to aquatic ecosystems and soil \cite{kent2002microbial,paerl1996mini}. In biotechnology, designing and controlling such communities is key to enabling sustainable production of biofuels, pharmaceuticals, and other valuable compounds \cite{diender2021synthetic,szotkowski2021bioreactor,mujtaba2017treatment}. Co-cultures comprising multiple microbial strains can outperform monocultures by leveraging division-of-labor strategies that reduce metabolic burden, improve pathway modularity, and enable processing of complex substrates \cite{bao2023engineering}.

Despite these advantages, industrial deployment of co-cultures remains limited, predominantly due to the lack of robust and scalable architectures for ensuring coexistence and maintaining stable consortium compositions. Without adequate control mechanisms, competitive exclusion can lead to community collapse \cite{chang2023emergent}. Control strategies are therefore essential to regulate growth and ensure long-term stability and productivity \cite{lee2022cybergenetic}.

Several works have explored feedback control strategies for microbial consortia through \textit{in silico} validation. For example, \cite{fiore2021feedback} investigated control of the population ratio between two species in a chemostat, proposing and comparing a gain-scheduled state feedback controller and a switching control law through simulation-based experiments. Similarly, \cite{asswad2024optimization} employed an optimization-based strategy to maximize algal biomass synthesis in a chemostat by controlling the dilution rate to address strain selection between two microbial species competing for substitutable substrates in a continuous stirred-tank reactor, where either species could potentially dominate the culture.

Recent works have demonstrated effective strategies for regulating microbial composition in compact bioreactors commonly used in laboratory settings. These strategies typically rely on genetic engineering to generate strains whose growth is tunable using external stimuli \cite{kusuda2021reactor,gutierrez2022dynamic,bertaux2022enhancing,lee2025directing,martinez2023optimal}. For example, Kusuda et al. \cite{kusuda2021reactor} created methionine- and arginine-auxotrophic E. coli strains, modulating their relative abundance by selectively feeding amino acids. By controlling nutrient inflow with simple feedback loops, this method avoids direct competition between strains, enabling stable co-culture. Alternatively, Gutierrez et al \cite{gutierrez2022dynamic} relied on light-induced antibiotic resistance to couple growth rate to external optical signals. In this setup, continuous antibiotic supply suppresses growth of the non-light-responsive strain, while blue light promotes survival of the engineered population. By dynamically tuning growth rates based on each strain's abundance, they created a synthetic fitness landscape where coexistence is actively managed through optogenetic control. Other studies have targeted metabolic coupling to enforce coexistence. For instance, Bertaux et al. \cite{bertaux2022enhancing} co-cultured yeast strains where one produced histidine, a metabolite essential for the other's survival. By adjusting nutrient availability through dilution rate control and employing model predictive control (MPC), they influenced composition. These strategies effectively balance the relative abundance of populations in microbial consortia. However, genetic interventions increase metabolic load on the host organism and can compromise production efficiency.

An alternative solution was presented in \cite{lee2025directing}, where a strategy to co-culture non-genetically modified strains of \textit{P. putida} and \textit{E. coli} was proposed. Here a single-chamber architecture that leverages the distinct temperature preferences of the two species was introduced. Specifically, changing culture temperature, the growth of either \textit{P. putida} or \textit{E. coli} was promoted.
The authors devised a feedback control law, coupled with an Extended Kalman Filter to estimate species ratios from indirect bioreactor measurements, which regulated consortium composition without genetic engineering. Although promising, this strategy requires dynamic temperature changes that can significantly alter microbial metabolism and interfere with bioproduction processes. Moreover, it requires the use of temperature-complementary strains, that is, strains whose relative growth rates can be reversed by changing the temperature of the co-culture. This assumption can be limiting, especially when efficient production requires specific growth conditions (e.g. \cite{zhou2015distributing}).

To overcome these limitations, we develop a general control architecture for managing naturally competing, non-complementary species without genetic modifications or dramatic changes in growth conditions. 
Specifically, we present a flexible and robust control architecture for regulating the composition and density of a two-strain consortium in real-time.
Building on our previous works \cite{brancato2024ratiometric,brancato2024vivo}, we constructed a bioreactor-based control architecture featuring two interconnected chambers: a mixing chamber hosting both strains, and a reservoir hosting a monoculture of the slower-growing strain. 
This setup enables dynamic modulation of growth conditions using three independent control inputs.
For each chamber, we designed a set of interchangeable control strategies leveraging the aggregate biomass measures that are automatically collected in the bioreactor. Specifically, we equipped the reservoir with a controller that can regulate the density of the slower-growing strain to a desired level, and the mixing chamber with a control law designed to stabilize the consortium at desired density and composition set-points.
For the control design, we investigated both model-based approaches and a learning-based controller developed through a sim-to-real pipeline, which leverages synthetic training data from a calibrated mathematical model.
The control architecture was validated \emph{in vivo} on a two strain \textit{E. Coli} consortium, demonstrating effective regulation of microbial consortium biomass and composition, including dynamic tracking of time-varying references and recovery from perturbations. Moreover, the learning-based approach achieved robust control with minimal experimental data, emphasizing the potential for data-efficient methods in synthetic biology applications.

\section{A control architecture for the composition control of two-strain bacterial consortia}
\label{sec:platform_and_problem}
\begin{figure}[!t]
    \centering
    \includegraphics[width=1\linewidth]{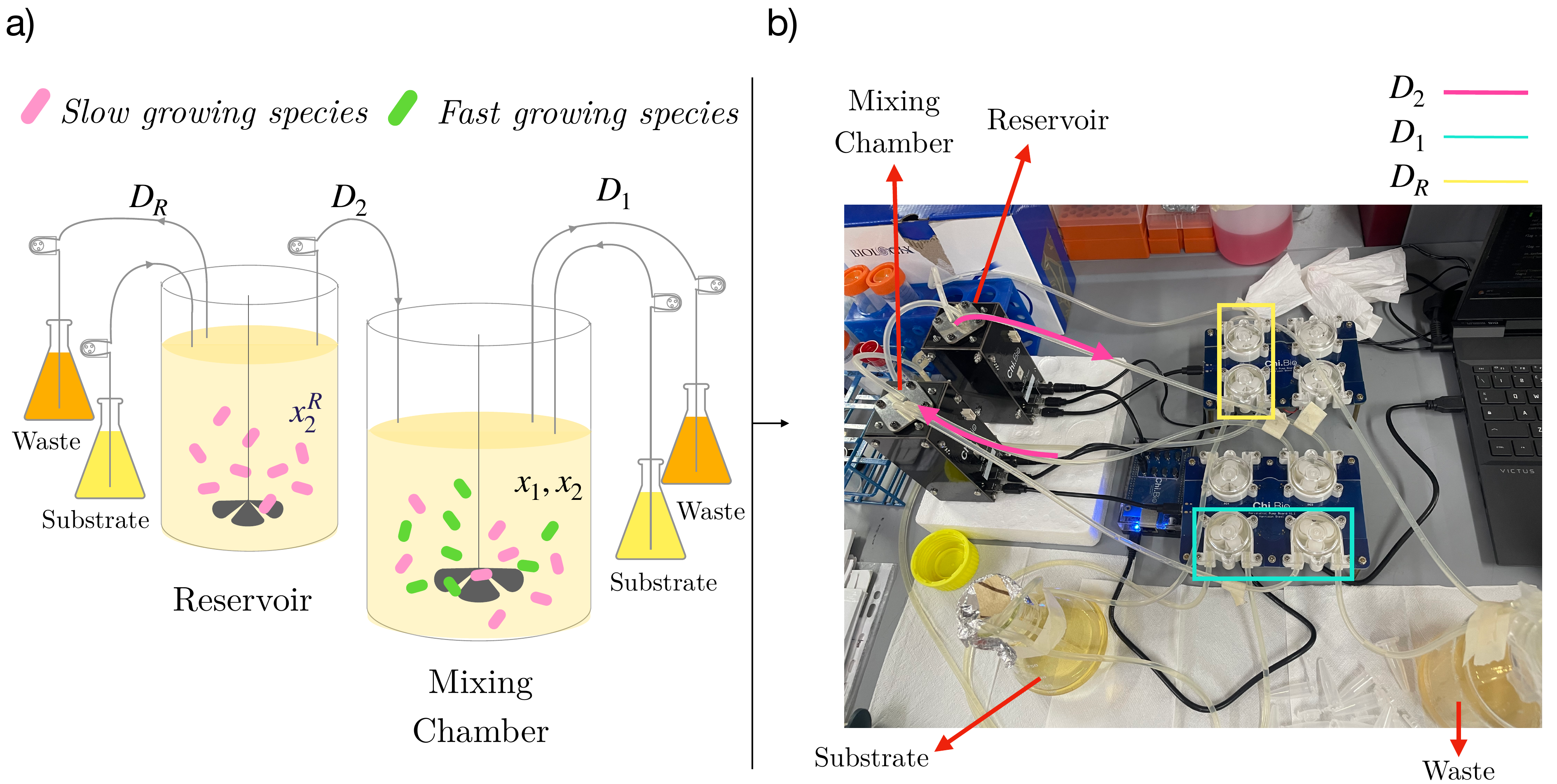}
    \caption{ a) A schematic representation of the control architecture, comprising a mixing chamber where strains 1 (in green) and 2 (in pink) are co-cultured, and a reservoir where the strain with a lower growth rate (i.e. strain 2) is grown independently. The mixing chamber and the reservoir are supplied with fresh growth medium at rates $D_1$ and $D_R$, respectively. Additionally, the reservoir feeds strain 2 to the mixing chamber at a rate $D_2$. Both reactors are connected to waste containers to ensure the volume is kept constant. b) Implementation of the control architecture with two interconnected Chi.Bio. reactors. The pumps that regulate the exchanges of fresh media and waste are highlighted with the teal and yellow boxes, while the tubes that interconnect the reservoir and the mixing chamber are highlighted with pink arrows.}
    \label{fig:architecture}
\end{figure}
We present a bioreactor-based control architecture designed for robust regulation of total density and ratio of a two-strain microbial consortium (see Figure \ref{fig:architecture}.a). Our control architecture controls two competing cell strains with non-complementary growth rates (i.e., one species always grows faster than the other).
The experimental setup comprises two interconnected reactor chambers. The first chamber, denoted as ``mixing chamber'', hosts a co-culture of both populations (strain 1 and strain 2), while the second chamber, named ``reservoir'', hosts a culture of the slower-growing population (strain 2). Both chambers are independently supplied with fresh media at rates that can be modulated in real time. Additionally, the mixing chamber can be fed with reservoir content at a controlled rate. Regulation is achieved using three control inputs: the flow rate of fresh media in the mixing chamber ($D_1$), the flow rate of the slower species from the reservoir into the mixing chamber ($D_2$), and the flow rate of fresh media in the reservoir ($D_R$).

\begin{figure}[!t]
    \centering
    \includegraphics[width=1\linewidth]{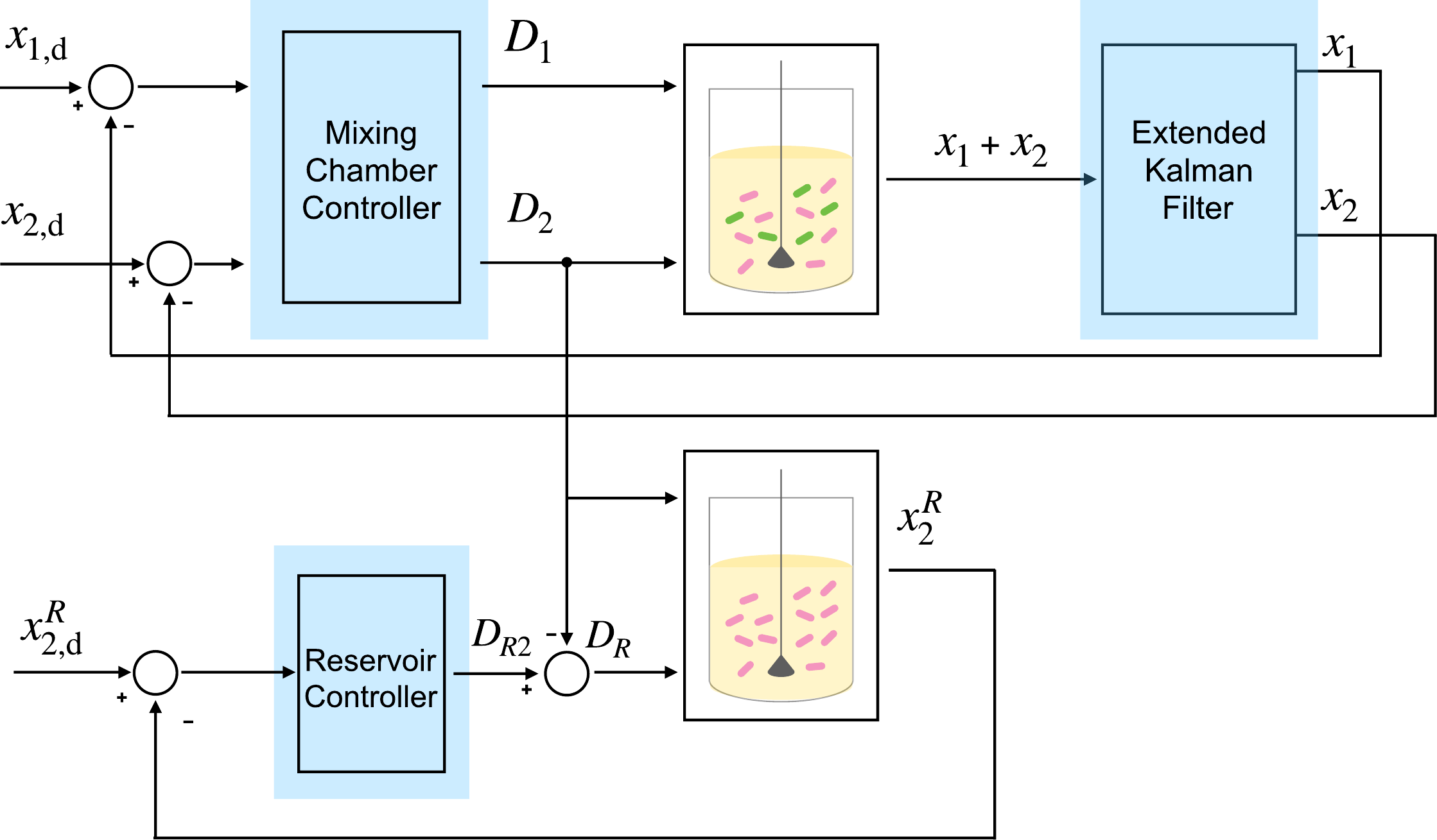}
    \caption{Block diagram of the closed-loop control architecture.
    The controller regulates the concentrations of the biomasses $x_1$ and $x_2$ in the mixing chamber to the desired levels $x_{1,\mathrm{d}}$ and $x_{2,\mathrm{d}}$, respectively. The current values of the states are reconstructed using an extended Kalman filter from the aggregate measure $x_1+x_2$. 
    An independently designed controller regulates $x_2^R$ in the reservoir to the desired level  $x_{2,\mathrm{d}}^R$.}
    \label{fig:block}
\end{figure}

The control algorithms endowed with the architecture are designed to coordinate the inputs to enable stable coexistence of both populations in the mixing chamber, with the ability to regulate density and composition at desired values. This is achieved by maintaining constant biomass concentration in the reservoir, which decouples reservoir biomass regulation from mixing chamber control. Specifically, by selecting control input $D_R$ to compensate for dilution rate $D_2$, we can independently regulate biomass density in the reservoir to a desired setpoint. The overall control architecture is illustrated in Figure \ref{fig:block}. Our objective is to design $D_1$ and $D_2$ to regulate total biomass ($x_1 + x_2$) and the ratio between populations ($x_1/x_2$) to desired levels in the mixing chamber, given the control architecture, cell populations, their growth dynamics, and available measures. Additionally, we aim to design a controller for $D_R$ to regulate biomass concentration $x_{2}^R$ in the reservoir to a desired reference setpoint. To avoid extinction events, we also require that $x_1$, $x_2$, and $x^R_2$ never fall below minimum density levels at all times.

To implement the control architecture, two Chi.Bio reactors \cite{steel2020situ} were interconnected and operated in coordination (Figure \ref{fig:architecture}.b). Each unit is equipped with a set of peristaltic pumps, all controlled by the onboard microcontroller, which also manages integrated sensors (e.g., optical density (OD), temperature). The OD measurement serves as an indicator of cell density, with data sampled every minute. The pumps are responsible for both supplying fresh media and removing waste in each chamber. Additionally, one dedicated pump transfers the slower-growing species from the reservoir into the mixing chamber.

While basic control algorithms can be implemented directly on the microcontroller, we developed a more flexible control framework based on a custom Python script running on a Windows PC. To send control inputs computed by the PC and read sensor measurements made by the onboard microcontroller, the two machines share a CSV file through an SFTP communication protocol. This setup allows the flexibility to design, fine tune, and deploy a wider spectrum of sophisticated control strategies, such as learning-based controllers, that may exceed the computational capabilities of the embedded system, while maintaining real-time interaction with the experimental platform. The sampling interval of one minute provides sufficient time for the connected PC to compute the control action and for the Chi.Bio system to apply the corresponding actuation.

\section{Model formulation and control goals}
To design our model-based control strategies, we derived a mathematical model based on the mass balance law, based on the approach used in \cite{bastin2013line}. For the reservoir chamber, we modeled the biomass and substrate dynamics as
\begin{equation}
\begin{cases}
\dot{x}^R_2 = \mu_2(s_2)x^R_2 - D_Rx^R_2 - D_2x^R_2,\\
\dot{s}_2 = -\mu_2(s_2)x^R_2 + (D_R + D_2)(s_{in} - s_2).
\end{cases}
  \label{eq:reservoir}
\end{equation}
Here, $x^R_2 \in \mathbb{R}_{\geq 0}$ and $s_2 \in \mathbb{R}_{\geq 0}$ denote the biomass concentration of the slow-growing strain (strain 2) and the feeding substrate concentration in the reservoir, respectively. Equation \ref{eq:reservoir} describes the balance between biomass created by the population, which consumes $s_2$ and grows at rate $\mu_2(s_2)$, and the dilution caused by fresh media inflow.

For the mixing chamber, we model the time evolution of both strains' biomass comprising the consortium, as well as the substrate concentration on which both species feed. The mass balance equations yield
\begin{equation}
\begin{cases}
\dot{x}_1 = \mu_1(s_1)x_1 - D_1x_1 - D_2x_1,\\
\dot{x}_2 = \mu_2(s_1)x_2 - D_1x_2 + D_2(x^R_2 - x_2),\\
\dot{s}_1 = - \mu_1(s_1)x_1 - \mu_2(s_1)x_2+\\
\quad\quad + D_1(s_{in} - s_1) + D_2(s_2 - s_1),
\end{cases}
\label{eq:mixing_chamber}
\end{equation}
where $x_1, x_2 \in \mathbb{R}_{\geq 0}$ denote the biomass concentrations of the fast-growing ($x_1$) and slow-growing ($x_2$) strains, respectively. Similarly, $s_1, s_2 \in \mathbb{R}_{\geq 0}$ are the substrate concentrations in the mixing chamber and reservoir, respectively. Also, $\mu_1(s_1), \mu_2(s_1)$ represent the growth rates of $x_1$ and $x_2$, respectively. Without loss of generality, we assume the yield coefficient for substrate-to-biomass conversion equals one.

The functions $D_1(t) : \mathbb{R}_{\geq 0} \mapsto [D_{1,\min}, D_{1,\max}]$, with $D_{1,\min} \geq 0$, and $D_R(t) : \mathbb{R}_{\geq 0} \mapsto [D_{R,\min}, D_{R,\max}]$, with $D_{R,\min} \geq 0$, represent our control inputs and are dilution rates defined as the inlet flow rate normalized by the reactor volumes, assumed equal for both chambers. Specifically, $D_1(t)$ and $D_R(t)$ dilute the mixing chamber and reservoir contents, respectively, by injecting fresh substrate at concentration $s_{in}$. In contrast, $D_2(t) : \mathbb{R}_{\geq 0} \mapsto [D_{2,\min}, D_{2,\max}]$ represents the media flow from the reservoir to the mixing chamber. Dilution effects on both strains are assumed identical, meaning the culture is well-mixed and bacterial mortality and attachment are neglected. Note that this is a standard assumption in the literature (see for example \cite{tani2019}). 

Growth rates $\mu_i(\cdot)$ are described by Monod law \cite{monod1949growth}, which is standard for cells feeding on non-toxic substrates. Namely,
\begin{equation}
\mu_i(s) := \mu^*_i \frac{s}{k_i + s}, \quad i = 1, 2,
\end{equation}
where $\mu^*_i > 0$ is the maximum growth rate of strain $i$ and $k_i > 0$ is the half-velocity constant. Assuming abundant substrate ($s \gg k_i$), the growth functions simplify to
\begin{equation}
\mu_i(s) \approx \mu^*_i, \quad i = 1, 2.
\label{eq:ass_mu}
\end{equation}
Note that this assumption holds at low cell densities, which are commonly used in bioreactors (see e.g. \cite{lee2025directing}).
Hence the model in equation \ref{eq:mixing_chamber} becomes:
\begin{equation}
\begin{cases}
\dot{x}_1 = \mu^*_1x_1 - D_1x_1 - D_2x_1\\
\dot{x}_2 = \mu^*_2x_2 - D_1x_2 + D_2(x^R_2 - x_2)
\end{cases}
\label{eq:simplified-model} 
\end{equation}
This assumption applies to the reservoir as well. By considering $\mu^*_1 > \mu^*_2$, we capture the heterogeneity in growth rates between the two bacterial populations and focus on the most challenging control scenario, where the populations in the mixing chamber would not coexist without proper regulation. In the mixing chamber we measure total biomass $y_1 = x_1+x_2$, while in the reservoir we measure $y_2 = x^R_2$.

\subsection{Control objectives}
Our overarching goal is to design $D_1$ and $D_2$ to regulate the total biomass ($x_1+x_2$) and the ratio between the two populations ($x_1 / x_2$) to desired levels in the mixing chamber, given the control architecture, the cell populations, their growth dynamics and the available measures $y_1$, $y_2$. Additionally, we aim to design $D_R$ to regulate the biomass concentration $x_2^R$ in the reservoir to a desired reference set-point.
To avoid extinction events, we also require that at all times $x_1$, $x_2$ and $x_2^R$ do not fall below a minimum density.

The control objectives can therefore be listed as follows:
\begin{enumerate}
\item \textit{Avoid extinction events:}
\begin{equation}
\begin{aligned}
x_1(t) &\geq x_{1,\min}, \quad \forall t > 0\\
x_2(t) &\geq x_{2,\min}, \quad \forall t > 0\\
x^R_2(t) &\geq x^R_{2,\min}, \quad \forall t > 0
\end{aligned}
\end{equation}
where $x_{1,\min}, x_{2,\min}, x^R_{2,\min}$ are positive densities;

\item \textit{Achieve total biomass regulation in the mixing chamber:}
\begin{equation}
\lim_{t \to \infty} \left(x_1(t) + x_2(t)\right)= OD_d,
\end{equation}
where $OD_d$ is the desired total biomass;

\item \textit{Regulate the ratio between populations in the {\em mixing chamber}:}
\begin{equation}
\lim_{t \to \infty} \left( x_2(t) - r_d x_1(t) \right) = 0,
\end{equation}
where $r_d$ encodes the desired consortium composition;

    \item \textit{Regulate biomass levels in the \emph{Reservoir}:}. \begin{equation}
        \lim_{t\to \infty} x^R_2(t) = x^R_{2,\mathrm{d}},
        \label{eq:desired_reservoir}
    \end{equation} 
    where $x^R_{2,\mathrm{d}}$ is the desired density of the slow-growing population in the reservoir.

\end{enumerate}

Note that simultaneously requiring total biomass and ratio regulation at desired values is equivalent to requiring
\begin{equation}
\begin{aligned}
\lim_{t \to \infty} x_1(t) &= \frac{1}{1 + r_d} OD_d\\
\lim_{t \to \infty} x_2(t) &= \frac{r_d}{1 + r_d} OD_d
\end{aligned}
\end{equation}

\section{Mixing chamber control design}\label{sec:mixing_chamber_control}
In this section, we present and analyze two control strategies designed to regulate both total density and species ratio within the mixing chamber: a learning-based control approach and a switching controller. 
To synthesize the controllers, we incorporate constraints from both biological considerations and technical limitations of the experimental setup.
To ensure safe operation of the Chi.Bio system, the maximum allowable dilution rate for all inputs is $0.02$ mL/s. The biomass is limited by the dilution rate, with an upper bound $x_{\max} = 1$ to ensure the Monod function approximation in \eqref{eq:ass_mu} holds. The $x_{\max}$ corresponds to a biomass concentration approximately 6–8 times lower than the maximum OD measured after overnight growth (i.e., at saturation). To prevent extinction, the dilution rate is never allowed to reduce biomass below the lower bound $x_{\min} = 0.2$.

Both controllers were designed under the following assumptions:

\begin{assumption} \label{ass:reservoir_concentration}
    The biomass concentration in the reservoir is constant and equal to $x^R_{2,\mathrm{d}}$.
\end{assumption}
\begin{assumption} \label{ass:direct_measure}
    Direct measurements of $x_1, x_2$ are available.
\end{assumption}

\begin{remark}
    Assumption~\ref{ass:reservoir_concentration} holds if the feedback controller in the reservoir chamber can stabilize the biomass of $x^R_2$ to the reference value $x^R_{2,\mathrm{d}}$ (see section \ref{sec:reservoir_control} for details on the reservoir control design).
\end{remark}

\begin{remark}
    Assumption~\ref{ass:direct_measure} will be addressed by designing a state observer based on the available measures (see section \ref{sec:Observer}).
\end{remark}

\subsection{Learning based control}

\begin{figure}[!t]
    \centering
    \includegraphics[width=0.7\linewidth]{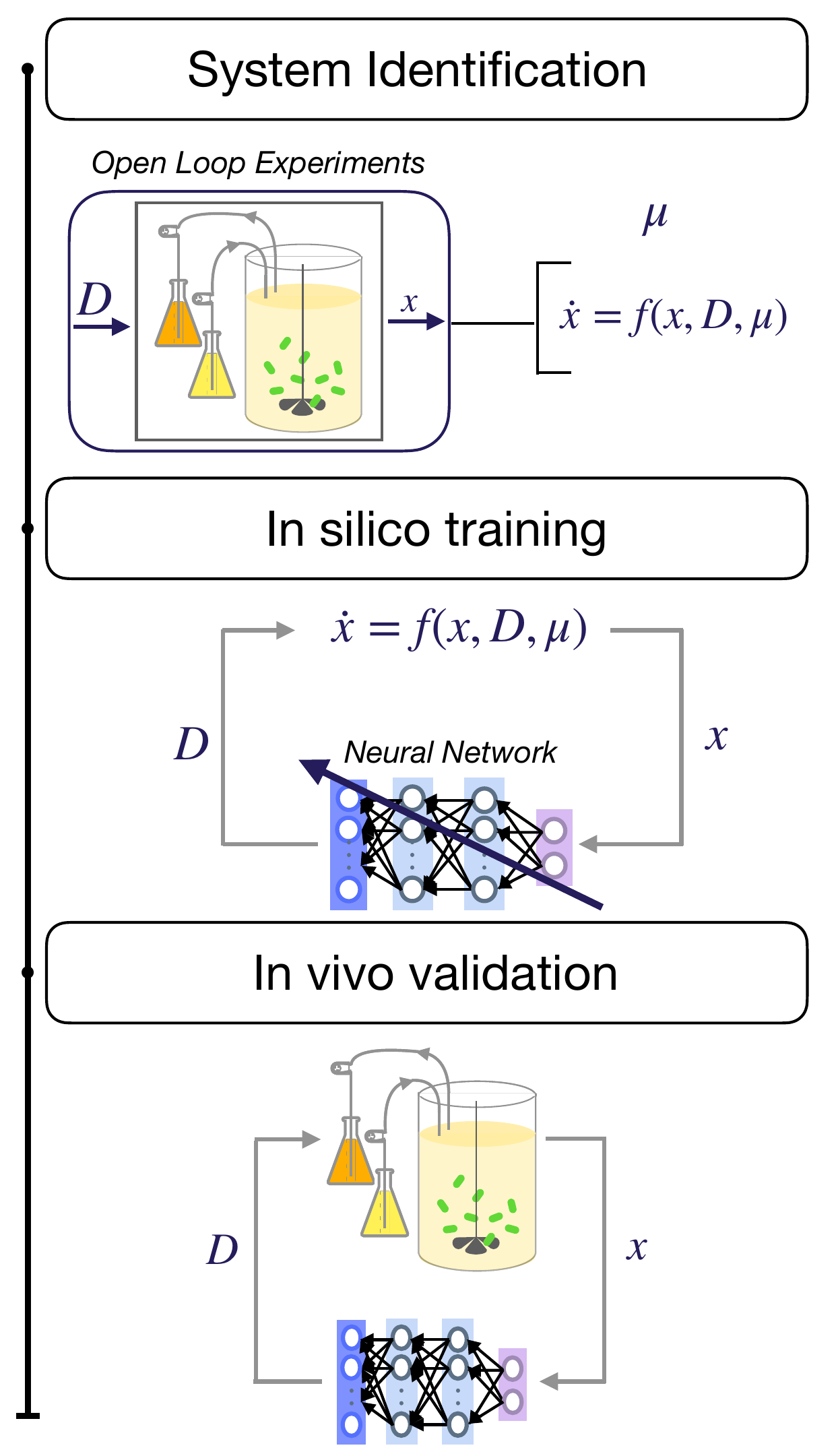}
    \caption{Sim-to-real pipeline. Step 1) System Identification: the parameters of the dynamical system are identified with an open-loop experiment;
    Step 2) \textit{In silico} training: the mathematical model is employed to produce synthetic data for the training of the neural network.  
    Step 3) \textit{In vivo} deployment:  The trained network is employed to regulate the cell population density within the bioreactor.}
    \label{fig:Figure3}
\end{figure}

We chose to implement a learning-based controller as this methodology provides near-optimal performance with the need of little information on the underlying dynamics of the system to be controlled. 
Specifically, the controller was designed following a sim-to-real approach, where synthetic data train a Reinforcement Learning agent (see Figure \ref{fig:Figure3} for a schematic description of the pipeline). The pipeline comprises three steps. First, we calibrated the dynamical model \eqref{eq:simplified-model} (see Section \ref{section:model_calibration}) with few open-loop experiments. Second, this mathematical model generates synthetic data to train a Reinforcement Learning agent. Third, the agent is deployed on the experimental platform, closing the gap between simulation and experiment. We implement a Deep Q-Network (DQN) algorithm, where a neural network approximates the optimal action-value function \cite{mnih2015human}. We chose this algorithm since we have a discrete set of available control actions and a continuous state space. 

The neural network receives as input a time series of the 5 latest concentration errors relative to desired values. Formally, the input vector is defined as
$\mathbf{x}_I = [x_1(t_{k-4})-x_{1,d}, x_2(t_{k-4})-x_{2,d}, \dots, \,  x_1(t_{k})-x_{1,d}, x_2(t_{k})-x_{2,d}],$ where $t_k = k \Delta t$ is the discretized time-step and $\Delta t$ is the discretization step. This choice results in 10 input features. The network architecture includes two hidden layers with 64 nodes each, and an output layer with 6 nodes corresponding to Q-values for all possible combinations of discrete control actions: $D_1 \in \{0.0, 0.01, 0.02\}$ and $D_2 \in \{0.0, 0.02\}$. These action sets were designed with few values to limit problem dimensionality and accommodate technological limitations of the peristaltic pumps, which cannot consistently deliver the same input at low actuation levels. The controller selects the action with the highest Q-value at each time step.

To improve generalization and enable regulation across various operating points, we randomized reference values at the start of each training episode. Specifically, $r_d$ was uniformly sampled from the discrete set $\{0.5, 1.0, 1.5\}$, while $OD_d$ was fixed at 0.7. Initial conditions for both state variables were drawn uniformly from the same sets. This strategy enables the DQN to learn a single policy capable of stabilizing the system around multiple setpoints, promoting versatility.

The reward function for training is defined as
\begin{equation}
R_t = r_{x_1} + r_{x_2} + P,
\end{equation}
where each $r_{x_i}$ penalizes deviations from the desired setpoint using the distance-like term:
\begin{equation}
r_{x_i} = -100 \tanh^2(w|x_i - x_{i,d}|),
\end{equation}
with parameter $w$ controlling tracking precision. A value of $w = 2$ was selected empirically as large enough to promote accurate tracking, but not so large as to hinder exploration by overly penalizing small deviations during training. To ensure biologically meaningful behaviors, we introduced the penalty term
\begin{equation}
P = \begin{cases}
-100 & \text{if } x_i < x_{i,\min}\\
0 & \text{otherwise,}
\end{cases}
\end{equation}
which discourages the controller from allowing either species to fall below a minimum viable concentration. \\
The DQN controller was trained using Google Colab on a single NVIDIA Tesla T4 GPU (16 GB VRAM, CUDA 12.2). Training required approximately 45 minutes for convergence over 200 episodes (180 samples each). During deployment, the inference time is on the order of milliseconds, allowing the controller to provide a control action well within the 1-minute sampling interval of the Chi.Bio measurement system.\\
To enhance robustness of the trained policy and improve generalization across different experimental conditions, we introduced both model uncertainty and measurement noise during training. For each episode, growth rates for both populations were sampled from Gaussian distributions centered at their nominal values $\mu^n_i$, which were previously identified (see Section \ref{section:model_calibration}). The sampling was defined as:
$\mu^*_i \sim \mathcal{N}(\mu^n_i, \sigma), \text{ with } \sigma = 0.15 \cdot \mu^n_i.$
This variability simulates uncertainty in system identification or biological variability across strains or conditions.

Additionally, we added Gaussian noise to the measurements provided to the neural network by adding $w_1(t) \sim \mathcal{N}(0, 0.001)$ and $w_2(t) \sim \mathcal{N}(0, 0.001)$ to $x_1$ and $x_2$ at each time step, respectively. These perturbations help the neural network learn policies that are more resilient to real-world disturbances, promoting reliable behavior when deployed on the experimental platform. The learned control policy is validated in Section \ref{section:validation_mixing}.

\subsection{Model-based switching control}

As an alternative to the learning-based controller, we developed a model-based switching controller adapted from previous work by some of the authors reported in \cite{brancato2024ratiometric}. 
This control algorithm was selected both for its simplicity and for the stability guarantees it is able to provide.
The switching controller relies on two switching surfaces that divide the state space $(x_1, x_2)$ into four regions. Choosing $\sigma_1 = x_1-x_{1,d}$ and $\sigma_2 = x_2-x_{2,d}$, the plane is divided into the following regions:
\begin{equation}
\begin{aligned}
\mathcal{R}_1 &= \{x_1, x_2 \in \mathbb{R}^+ \mid x_1 < x_{1,d} \land x_2 > x_{2,d}\},\\
\mathcal{R}_2 &= \{x_1, x_2 \in \mathbb{R}^+ \mid x_1 > x_{1,d} \land x_2 > x_{2,d}\},\\
\mathcal{R}_3 &= \{x_1, x_2 \in \mathbb{R}^+ \mid x_1 < x_{1,d} \land x_2 < x_{2,d}\},\\
\mathcal{R}_4 &= \{x_1, x_2 \in \mathbb{R}^+ \mid x_1 > x_{1,d} \land x_2 < x_{2,d}\},
\end{aligned}
\end{equation}

The control inputs are designed as
\begin{equation}
(D_1, D_2) = \begin{cases}
(D^+_1, 0) & \text{if } (x_1, x_2) \in \mathcal{R}_1\\
(D^-_1, 0) & \text{if } (x_1, x_2) \in \mathcal{R}_2\\
(0, 0) & \text{if } (x_1, x_2) \in \mathcal{R}_3\\
(0, D^-_1) & \text{if } (x_1, x_2) \in \mathcal{R}_4
\end{cases}
\end{equation}
Setting $D^+_1 > \mu_1(s_{in})$ and $D^-_2 > \mu_1(s_{in})$, the system asymptotically reaches the desired equilibrium $(x_{1,d}, x_{2,d})$ (see \cite{brancato2024ratiometric} for the analytical proof of convergence). Note that for $(D_1, D_2) = (D^+_1, 0)$ when $x \in \mathcal{R}_1$, stability of the solution $x_d$ is not affected because the trajectory still reaches one of the switching manifolds. The switching control policy is validated in Section \ref{section:validation_mixing}.

\subsection{Observer Design} \label{sec:Observer}
Since species-specific concentrations in the mixing chamber cannot be directly measured, we implemented a state observer to estimate $x_1$ and $x_2$ from the total biomass measurement $y_1 = x_1 + x_2$. We first verified system observability (see Appendix A for details) and then employed an Extended Kalman Filter (EKF) with noise covariances and initial uncertainty tuned to minimize estimation error. 

\section{Reservoir control design}
\label{sec:reservoir_control}

The assumption of constant $x^R_2$ made in Section \ref{sec:mixing_chamber_control} requires a controller that robustly regulates $x^R_2$ to the desired value. To achieve this, we explored three interchangeable control strategies: a PI controller already integrated into the Chi.Bio bioreactors, a model-based MPC, and a learning-based controller developed using a sim-to-real approach.
We selected the PI controller because it was already integrated into the Chi.Bio bioreactors. Instead, the MPC was included to evaluate the performance of optimal controllers based on accurate knowledge of the system model. Finally, a learning-based approach was employed to examine the optimality of methods that require minimal prior knowledge of the system dynamics.

As stated in Section \ref{sec:platform_and_problem}, our aim is to regulate biomass concentration $x^R_2$ to any desired steady-state value $x^R_{2,d}$ by choosing appropriate control input $D_R$, given measurement $y_2$. The constraints introduced for the mixing chamber also apply to the reservoir: the dilution rate is bounded by $D^{\max}_R = 0.02$ mL/s to prevent overflow, and biomass concentration is maintained within $[0.2, 1]$ through control dilutions. We explored multiple controllers to compare their performance and robustness under identical experimental conditions.

To simplify the control task, we decouple regulation of $x^R_2$ from regulation of total biomass and ratio in the mixing chamber through disturbance compensation. Specifically, dilution rate $D_R$ is chosen as $D_R = D_{R2} - D_2$, where $D_{R2}$ is the dilution rate designed for biomass regulation, and $D_2$ represents a measurable disturbance component known and determined by the mixing chamber controller (see Figure \ref{fig:architecture}). Under assumption \eqref{eq:ass_mu} , we can recast the reservoir dynamics from equation \eqref{eq:reservoir} as
\begin{equation}
\dot{x}^R_2 = \mu^*_2 x^R_2 - D_{R2} x^R_2.
\label{eq:reservoir_D12} 
\end{equation}
For simplicity, we designed all proposed controllers setting $D_2 = 0, \forall t$ (i.e., $D_{R2} = D_R$).
The PI controller uses the same structure and parameters described in \cite{steel2019chi}. The MPC uses model \eqref{eq:reservoir_D12} to minimize at every time step the cost function:
\begin{equation}
J(t) = \sum_{k=0}^{N-1}{ c(t+k\Delta t) + V_F(x(t+N\Delta t))},
\end{equation}
where $N$ is the prediction horizon, $ \Delta t = 1 \, $min is the sampling time, $c(\cdot)$ is  the cost, and $V_F(x(t+N\Delta t))$ is the final cost. 
We choose $N=5$ as a trade off between computational burden and control performance. Additionally, we chose cost function as  
\begin{equation}
c(t) = \begin{cases}100 & \quad  \text{if }  u(t) \notin [0,0.02]\\
(x_2^R(t)-x_{2,\mathrm{d}}^R)^2 & \quad  \text{otherwise }
\end{cases} ,
\end{equation}
and $V_F(x) = (x_2^R(t)-x_{2,\mathrm{d}}^R)^2$.
This formulation is designed to penalize both the deviation from the desired density and any violation of the constraints on the actuators. 

Finally, the learning-based controller was designed following the sim-to-real approach described in the previous section. Unlike the mixing chamber control, the Q-Network policy for reservoir control receives current optical density measurement $x^R_2(t)$ and desired regulation value $x^R_{2,d}$ as additional input features to improve generalizability across different reference values \cite{7983780}. Training data are synthetically generated using the simplified model \eqref{eq:reservoir_D12}. During training, the reference is randomly sampled from the discrete set $\{0.2, 0.3, \ldots, 0.9, 1\}$ at the start of each episode, enabling the agent to regulate the system across multiple setpoints using a single Q-network. Initial conditions are similarly drawn from the same set. The action space consists of 17 discrete values uniformly distributed in $[0, 0.02]$ for control input $D_{R2}$. The network architecture comprises two fully connected layers with 64 neurons each, activated by ReLU functions, and is trained using the Adam optimizer with learning rate 0.001. The reward function minimizes the control error and is set as:
\begin{equation}
r(t) = -(x^R_2(t) - x^R_{2,d})^2.
\end{equation}
The trained network was deployed \emph{in vivo} to control $x^R_2$ in the reservoir.

The three controllers represent different trade-offs between implementation complexity, computational requirements, and theoretical guarantees. The PI controller offers simplicity and is already integrated into the Chi.Bio bioreactors, the MPC provides model-based optimization with constraint handling, while the learning-based controller offers flexibility and can adapt to various setpoints without redesign. The performance and robustness of these controllers are experimentally compared in the following section.

\section{In Vivo Validation}
In this section, we present experimental validation of the proposed control framework. We begin by identifying model parameters through open-loop experiments. Since only total biomass is measurable in the mixing chamber, we implement a state observer to estimate individual species concentrations. We then define performance metrics for quantitative evaluation. The reservoir controllers are independently assessed and compared. Finally, we validate the complete control architecture, with both reservoir and mixing chamber controllers operating simultaneously.
Each experiment was performed in triplicate to balance experimental effort with statistical relevance.

\subsection{Model identification}
\label{section:model_calibration}

We selected two \textit{E. coli} strains as case study corresponding to two implementations of the genetic toggle switch presented in \cite{litcofsky2012iterative} (fast growing strain) and \cite{lugagne2017balancing} (slow growing strain), respectively.
To identify parameters in equations \eqref{eq:simplified-model}-\eqref{eq:reservoir_D12}, we conducted two open-loop experiments, growing each bacterial strain independently in the mixing chamber  (setting either $x_1(0)=0$ or $x_2(0)=0$, with $D_2=0$). 
Cultures were grown without dilution ($D_1 = 0$) until measured OD reached 1.0, then diluted, after which the dilution rate was randomly changed every 30 minutes.

\begin{figure}
    \centering
    \includegraphics[width=1\linewidth]{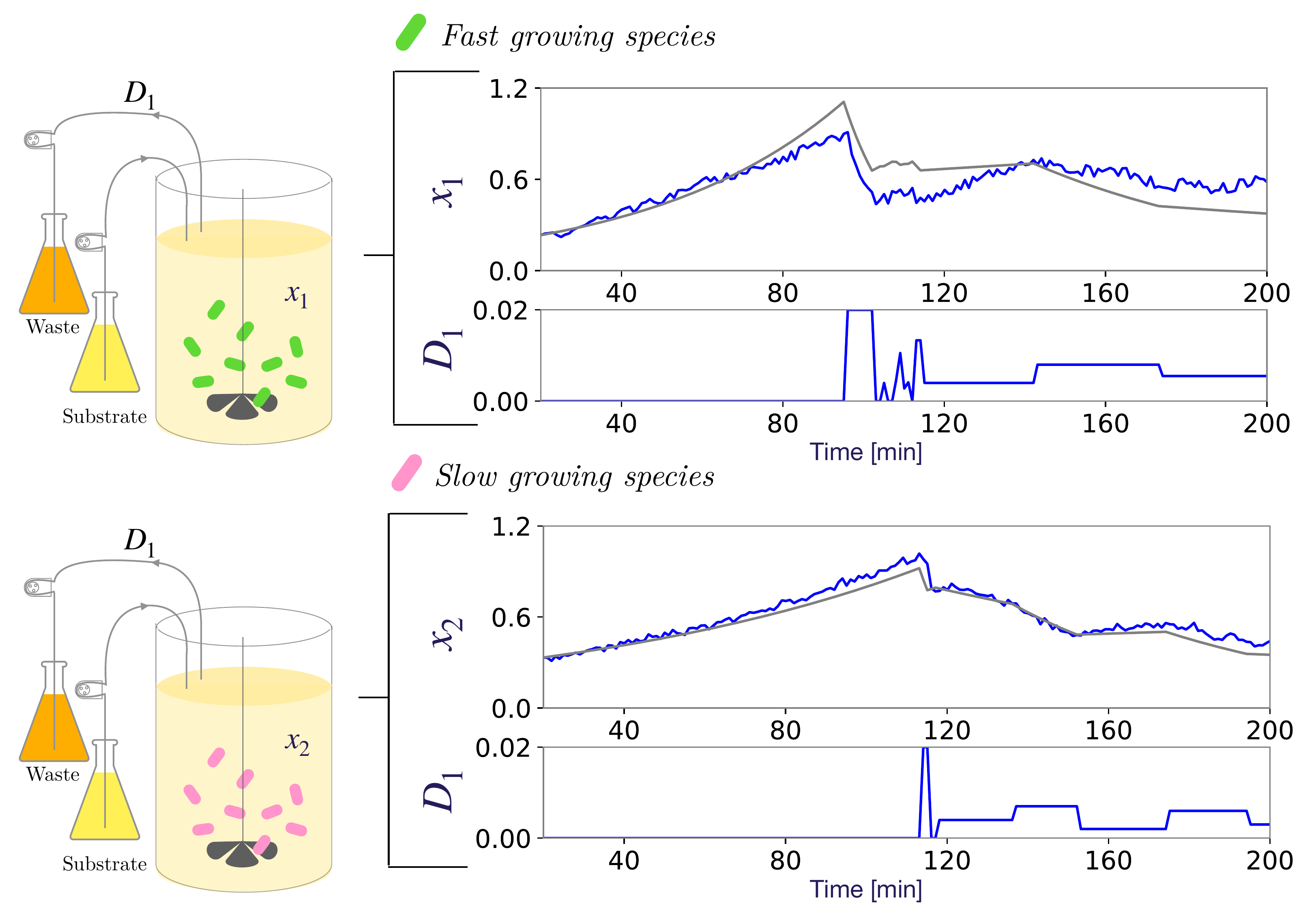}
    \caption{ Open loop experiments to identify the parameters $\mu_1^*$ and $\mu_2^*$. Time evolution $y_1$ when growing fast-growing strain (top) and slow-growing strain (bottom)  monocultures in the mixing chamber, while imposing $D_2=0$. A qualitative comparison between the experimental data (in blue) and the model predictions (in grey) is shown. For each experiment the dilution rate $D_1$ administered during the experiment is shown. }
    \label{fig:Identification}
\end{figure}

Experiments were performed at $37$ °C in Luria broth supplemented $1 mM$ IPTG (Isopropyl $\beta$-D-1-thiogalactopyranoside) to ensure that both populations expressed two different fluorescent reporters at high levels.
Using least-squares estimation in MATLAB on the resulting time series, we estimated growth rates $\mu^*_1 = 0.021$ and $\mu^*_2 = 0.011$. Additionally, we identified a scaling parameter $\tau = 0.215$ that captures dilution effects on system dynamics; each control input $D_1$, $D_2$, and $D_R$ is divided by this factor. 
Figure \ref{fig:Identification} compares simulated dynamics (gray) with experimental data (blue), showing good qualitative agreement. Root mean squared errors between simulations and experiments were 3.83 for $x_1$ and 0.48 for $x_2$.

\subsection{Observer Validation} \label{sec:Observer_validation} 

\begin{figure}[!t]
    \centering
    \includegraphics[width=1.0\linewidth]{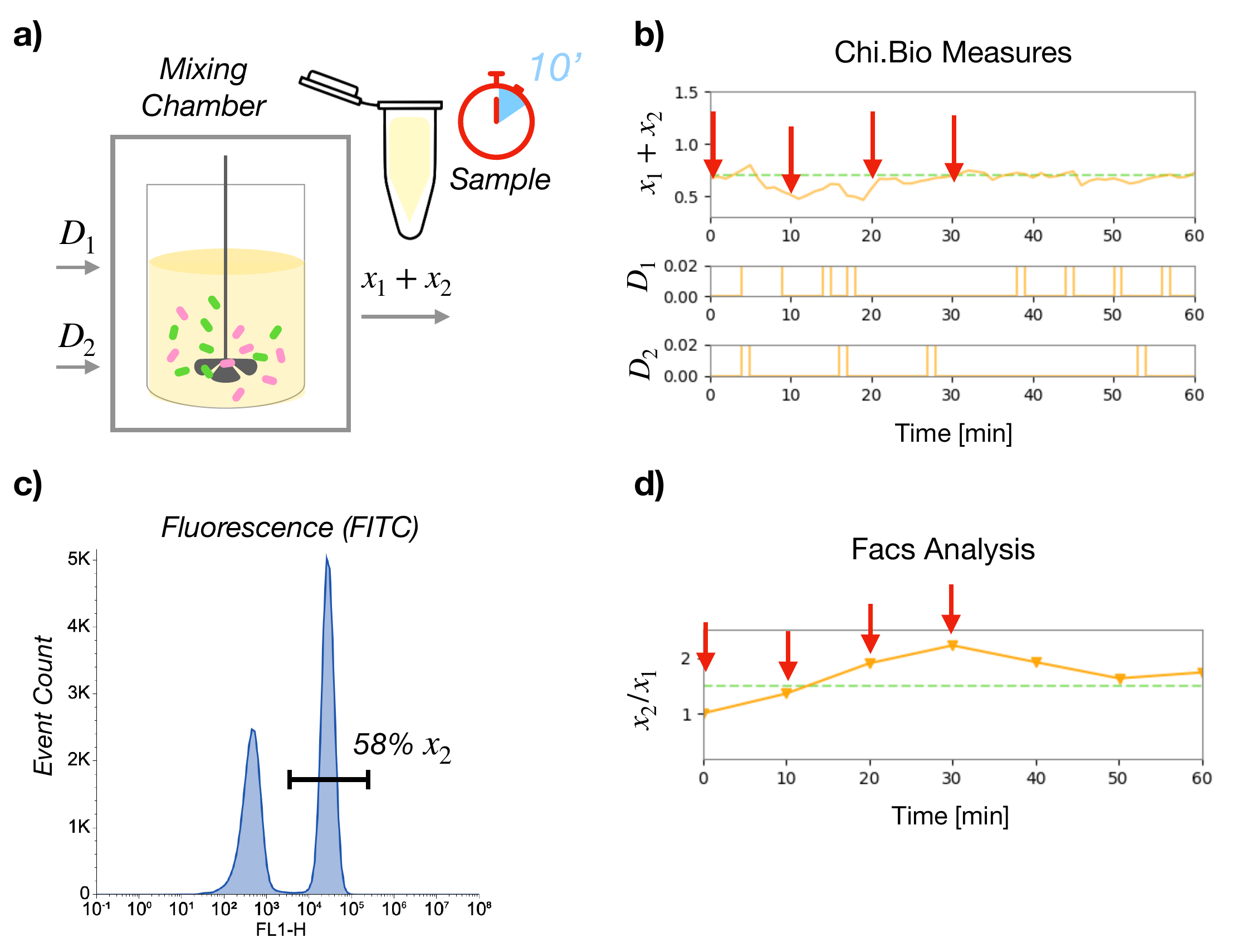}
    \caption{Experimental protocol to measure ex-post the relative numbers between the strains comprising the consortium in the mixing chamber. a) Every 10 minutes, samples are taken from the mixing chamber. b) Example measurements collected from the Chi.Bio. Top panel: time evolution of $y_1$. The red arrows indicate the first four time instants in which the samples are collected; bottom panel: dilution rates. c) Gating strategy used to distinguish the strains using a flow cytometer. After excluding debris and cell agglomerates using the FSC and SSC channels, cells were classified by their fluorochrome (FITC). Specifically, cells expressing high levels of green fluorescent protein (FL1-H $\geq 2\times 10^3$ on log scale), were classified as belonging to strain 2. The remaining cells were classified as belonging to strain 1. 
    d) Example of $x_2/x_1$ ratio computed using the data classified using the flow cytometer.
    }
    \label{fig:FACS}
\end{figure}

\begin{figure}[!t]
    \centering
    \includegraphics[width=1.0\linewidth]{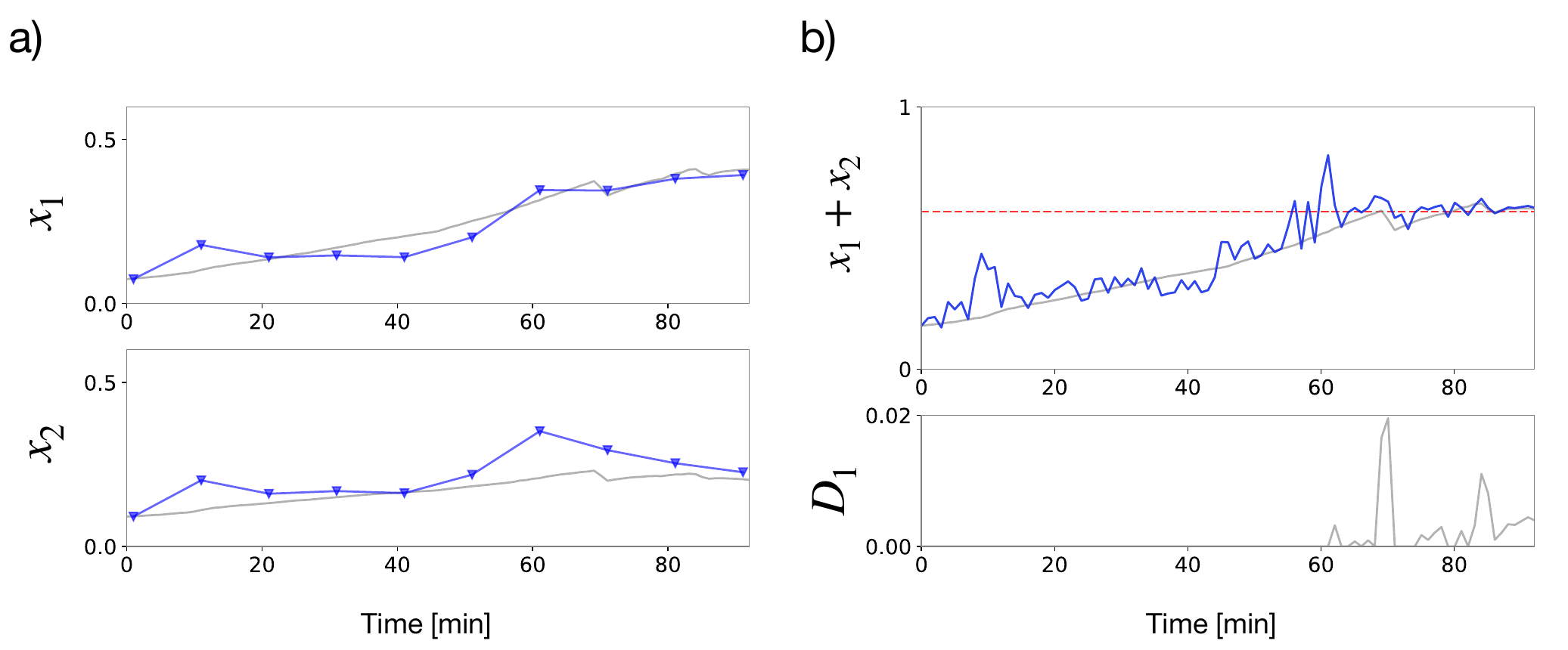}
    \caption{Consortium density regulation experiment ($D_2 = 0$). a) in gray the KF estimates of $x_1$,$x_2$, and in blue the measurement density of each population reconstructed using the flow cytometer. b) Top panel, In blue, $y_1$ measurement coming from the Chi.Bio. In gray the reconstruction of the sum $x_1 +x_2$ by KF estimates. Bottom panel: dilution rate $D_1$, is turned off for 60 min and then it regulates the OD to 0.6.}
    \label{fig:Validation1}
\end{figure}

To validate the observer, we co-cultured both populations in the mixing chamber at initial density $x_1 = x_2 = 0.1$ and regulated total biomass to $y_1 = 0.6$ using $D_1$ while setting $D_2 = 0$. Every 10 minutes, we sampled the culture and measured individual species concentrations using flow cytometry, distinguishing strains by fluorescence (strain 2 expresses GFP at high levels when induced with 0.1 mM IPTG, while strain 1 does not). Figure \ref{fig:FACS} illustrates the experimental protocol. Figure \ref{fig:Validation1} compares flow cytometry measurements with EKF estimates, showing good qualitative agreement with mean squared errors of 0.001 for $x_1$ and 0.004 for $x_2$. 

This experiment also demonstrates that regulating only total biomass causes $x_1$ to increase relative to $x_2$, potentially leading to extinction of the slower species, highlighting the importance of controlling both density and composition of the consortium.


\subsection{Control Metrics}
\label{sec:metrics}
To quantitatively compare control strategies, we use two metrics: settling time and normalized root mean squared error (NRMSE) at steady state. The settling time $t_s$ is the earliest time at which the system output enters and remains within a tolerance band of $\pm20\%$ around its steady-state value. The steady-state value is estimated as the average output over the final $20\%$ of the experiment. The NRMSE between measured variable $x$ and desired reference $x_d$ is computed as
\begin{equation}
NRMSE = \frac{1}{T - t_s} \int_{t_s}^T \sqrt{\frac{(x(t) - x_d)^2}{x_d}} dt
\end{equation}
where $T$ is the experiment duration.
The tolerance bound accounts for both the noise inherent in the Chi.Bio bioreactors and uncertainty introduced by flow cytometry (FACS) measurements used for ex-post ratio quantification. Based on similar experimental studies conducted on the same bioreactors \cite{lee2025directing}, a deviation of this magnitude can be considered an adequate threshold for defining successful regulation \emph{in vivo}.
\subsection{Reservoir}

\begin{figure}
    \centering
    \includegraphics[width=1.0\linewidth]{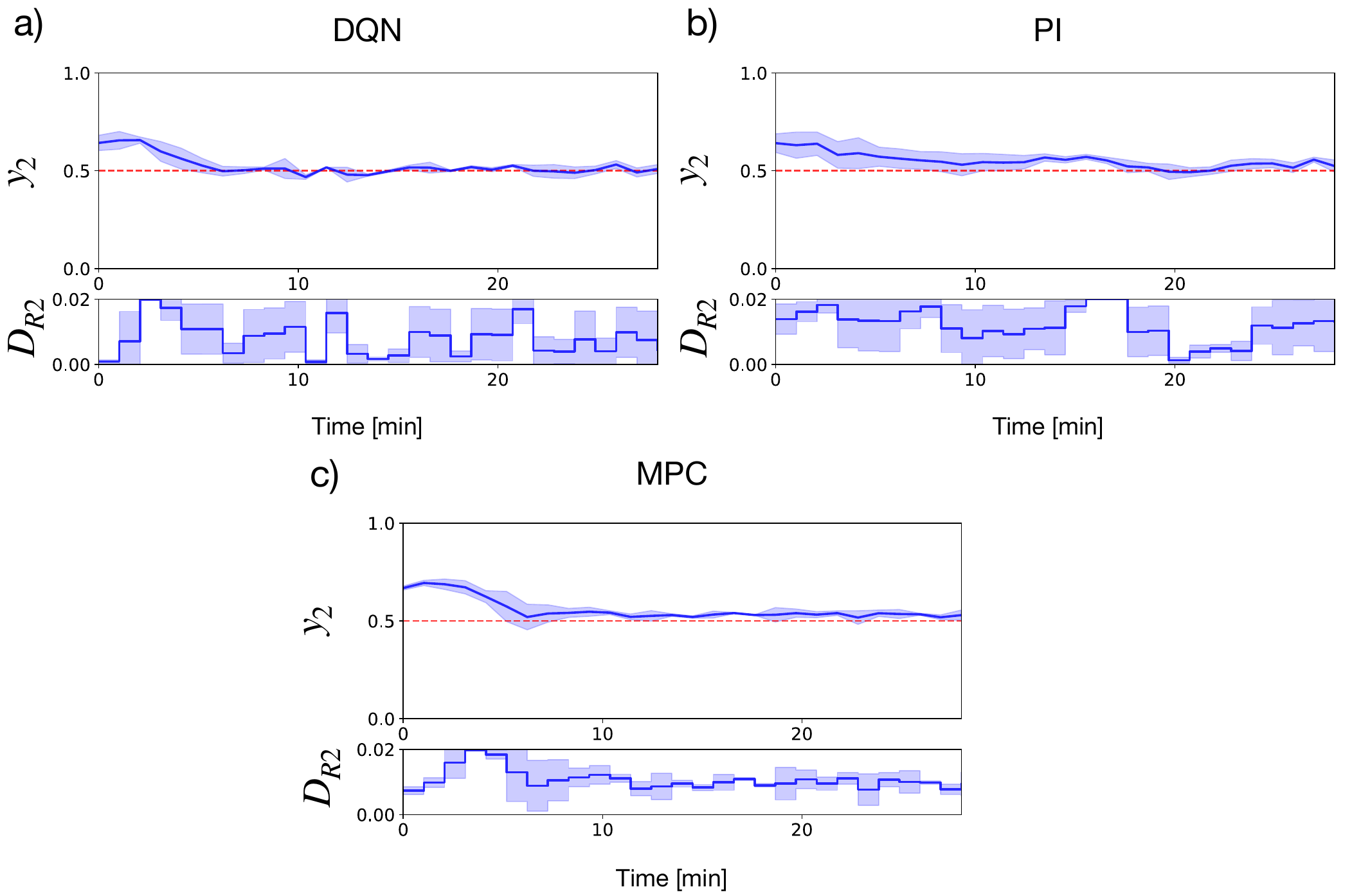}
    \caption{Comparison of the controllers performance in regulating $x_2^R$ to a desired value. For all experiments $x_{2,\mathrm{d}}^R = 0.5$. The top panels depict the time evolution of $y_2$ while the bottom panels depict the control input $D_{2R}$ computed by a) DQN, b) PI, c) MPC, respectively.
    Solid lines (in blue) represent the average evolution of state and input over the three \textit{in vivo} experiments, the light blue shadow represents the standard deviation. The red dashed lines indicate the desired set point.}
    \label{fig:Single_Population}
\end{figure}

We tested all three controllers designed in section \ref{sec:reservoir_control} experimentally. For each experiment, overnight cultures were transferred to the reservoir and allowed to recover for one hour at 37°C in LB. Cultures were then diluted to initial biomass concentration 0.8, which was maintained as the reference value for 30 minutes before stepping down to $x^R_{2,d} = 0.65$ for 30 minutes, then to $x^R_{2,d} = 0.5$ for the final 30 minutes. Figure \ref{fig:Single_Population} shows $y_2$ evolution under each controller. All controllers consistently stabilized biomass at different setpoints, with settling times under 7 minutes and NRMSE under 5\% (Table \ref{table:1}), demonstrating comparable performance across all three approaches.

\begin{table}[h!]
\centering
\caption{Average Settling Time and NRMSE Across Experiments (Mean ± Std)\label{table:1}}
\begin{tabular}{lcc}
\toprule
\textbf{Controller} & \textbf{Settling Time (min)} & \textbf{NRMSE} \\
\midrule
PI   & $5.01$  & $0.042$ \\
MPC  & $5.36$  & $0.047$ \\
DQN  & $6.91$  & $0.036$ \\
\bottomrule
\end{tabular}
\end{table}

\begin{table}[h!]
\centering
\caption{NRMSE Before and After Temperature Change \label{table:2} }
\begin{tabular}{lcc}
\toprule
\textbf{Controller} & \textbf{NRMSE (37$^\circ$C)} & \textbf{NRMSE (30$^\circ$C)} \\
\midrule
PI   & $0.027$ & $0.024$ \\
MPC  & $0.029$ & $0.028$ \\
DQN  & $0.032$ & $0.025$ \\
\bottomrule
\end{tabular}
\end{table}

To assess robustness, we tested the controllers under temperature perturbations that affect cellular growth rate. We experimentally observed that decreasing temperature from 37°C to 30°C reduces growth rate by approximately 10\%. Time course experiments maintained temperature at 37°C for 30 minutes, then decreased it to 30°C for 30 minutes. All controllers successfully maintained biomass concentration at the desired value despite the growth rate perturbation. Table \ref{table:2} shows that NRMSE values remained comparable before and after the temperature change, confirming robustness of all three controllers.

Given comparable performance across all three controllers, we selected the embedded PI controller for subsequent experiments due to its simplicity and direct integration with the Chi.Bio bioreactor, requiring no additional computational infrastructure. For integration with the mixing chamber control, we set $D_R = D_{R2} - D_2$, where $D_2$ is determined by the mixing chamber controller (Figure \ref{fig:block}).

\subsection{Mixing Chamber}
\label{section:validation_mixing}

For mixing chamber control, we coupled the observer with either the model-based switching controller or the learning-based controller. The reservoir used the PI controller from Section \ref{sec:reservoir_control} with $D_{R2}$ as described there. To prevent depletion of the slower species, we implemented a recovery mechanism: when reservoir concentration falls below 0.8, $D_2$ is set to zero until concentration recovers above 0.8.

We validated both controllers through three experimental scenarios: regulation to fixed setpoints, tracking of time-varying references, and recovery from perturbations. All experiments used the control architecture from Section \ref{sec:platform_and_problem}. Before each experiment, overnight cultures were diluted 1:3 and grown independently in Chi.Bio reactors for 1 hour. The mixing chamber was then filled with both populations at defined ratio and density, while the reservoir was initialized with the slow-growing population at $x^R_2(0) = 0.8$.

\begin{figure}[!t]
    \centering
    \includegraphics[width=1\linewidth]{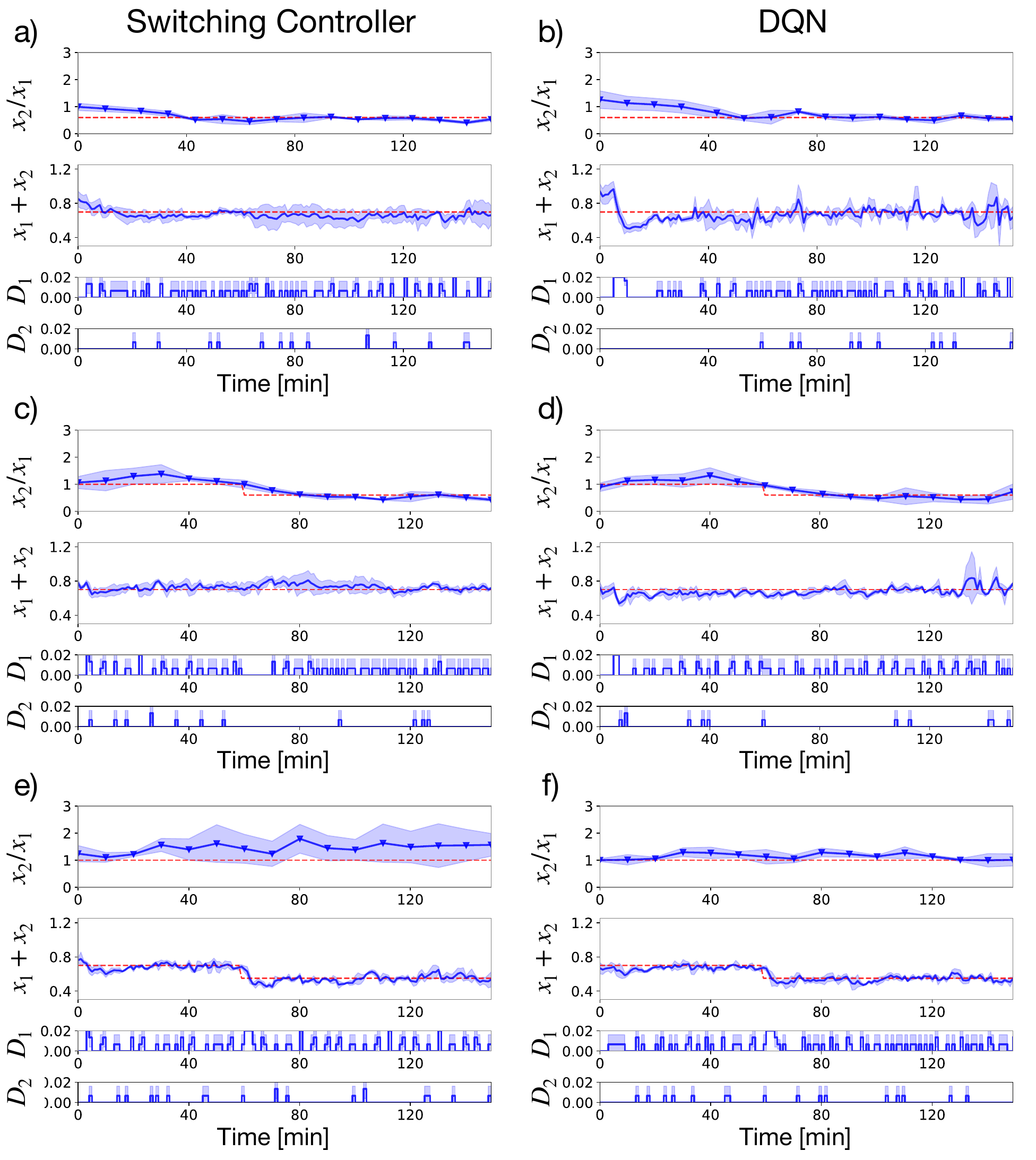}
    \caption{Experimental validation of the regulation and tracking capabilities of the control algorithms across three different scenarios. Each panel displays mean (solid line) and standard deviation (shaded areas) computed over three biological replicates. In each scenario, it is shown the evolution of the ratio, the total biomass and the control inputs in blue. Additionally, the reference points are represented as dashed red lines.  a)-b) Regulation: Starting from $(x_1(0),x_2(0))=(0.4,0.4)$ we set $r_d=0.6$, $OD_{\mathrm{d}}=0.7$. c)-d)Tracking of two constant ratio's set points:
    Starting from an initial condition of $(x_1(0),x_2(0))=(0.35,0.35)$ we set $r_d=1$, $OD_{\mathrm{d}} =0.7$ for an hour, then changed $r_d = 0.6$. e)-f)Tracking of two constant total biomass set points: Starting from $(x_1(0),x_2(0))=(0.35,0.35)$, we set $r_d=1$, $OD_{d} =0.7$ for an hour, then changed $OD_{d} = 0.55$.}
    \label{fig:composition_validation}
\end{figure}

\paragraph{Regulation to fixed setpoints} Starting from $x_1(0) = x_2(0) = 0.4$, we set reference values $r_d = 0.6$ and $OD_d = 0.7$. Samples collected every 10 minutes and analyzed by flow cytometry (protocol in Figure \ref{fig:FACS}) confirmed that both controllers steered the consortium toward desired setpoints (Figure \ref{fig:composition_validation}a,b). Table \ref{table:mixing_performance} shows both controllers achieved desired composition and density within one hour, with steady-state errors under 10\% for both ratio and total biomass. The higher error in ratio regulation is primarily attributable to Kalman Filter reconstruction; computing the error metric on observer-reconstructed trajectories yields errors under 1\%.
To quantitatively assess whether the two controllers exhibited statistically different performances, we performed a paired two-tailed t-test comparing the settling times and steady-state NRMSE values obtained across the three biological replicates for each experimental scenario. The resulting p-values (reported in Table\ref{table:mixing_performance}) were all greater than 0.1, indicating no statistically significant difference between the switching controller and the DQN controller. 

\paragraph{Tracking time-varying references} To assess real-time setpoint adaptation, we performed two tracking experiments starting from $(x_1(0), x_2(0)) = (0.35, 0.35)$. In the first experiment, we changed composition while maintaining constant total biomass: $OD_d = 0.7$ and $r_d = 1$ for one hour, then $r_d = 0.6$ for the remainder. In the second experiment, we changed total biomass while maintaining constant composition: $r_d = 1$ throughout, with $OD_d = 0.7$ for one hour, then $OD_d = 0.55$. Both controllers successfully tracked the changing setpoints (Figure \ref{fig:composition_validation}c-f). Table \ref{table:mixing_performance} shows faster settling when one variable remains constant, with average errors around 20\% for ratio and 10\% for total biomass.

\begin{figure}[!t]
    \centering
    \includegraphics[width=1\linewidth]{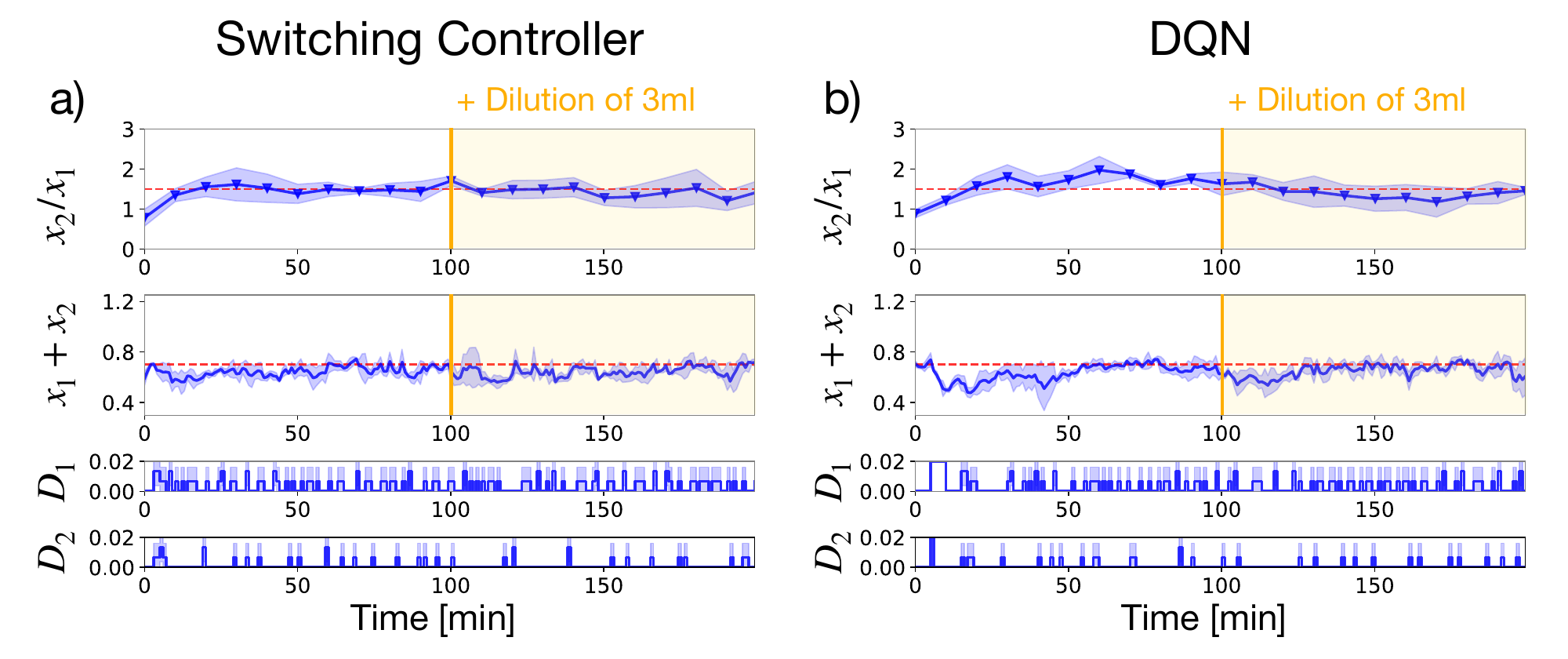}
    \caption{Experimental demonstration of the robustness capabilities of the control architecture to external perturbations.
    Starting from $(x_1(0),x_2(0))=(0.35,0.35)$, we set $r_d=1.5$, $OD_d =0.7$. After 100 minutes 3 mL of fresh media were added to the mixing chamber. Each panel displays mean (solid line) and standard deviation (shaded ares) computed over three biological replicates. For both controllers it is shown the evolution of the ratio, the total biomass and the control inputs in blue. Additionally, the reference points are represented as dashed red lines. The shaded yellow area highlights the time interval following the introduction of the external disturbance.} 
    \label{fig:robustness}
\end{figure}

\paragraph{Robustness to perturbations} To assess disturbance rejection, we started from $(x_1(0), x_2(0)) = (0.35, 0.35)$ with setpoints $r_d = 1.5$ and $OD_d = 0.7$. After 100 minutes, we injected 3 mL of fresh media into the mixing chamber, perturbing both total biomass and composition. Both controllers recovered from the disturbance after a short transient, with maximum ratio deviation within 20\% (Figure \ref{fig:robustness}). Table \ref{table:mixing_robustness} shows that ratio regulation performance remained comparable before and after perturbation, confirming controller robustness.

\begin{table*}[ht]
\centering
\caption{Comparison between Switching (S.C.) and DQN controllers across the three experimental scenarios. Settling time (TS) and normalized root mean square error (NRMSE) are reported together with paired two-tailed \textit{t}-test $p$-values.}
\label{table:mixing_performance}
\begin{tabular}{l l
                c c c
                c c c}
\toprule
\multirow{2}{*}{\textbf{Experiment}} & \multirow{2}{*}{\textbf{Signal}} &
\multicolumn{3}{c}{\textbf{Settling Time (TS)}} &
\multicolumn{3}{c}{\textbf{NRMSE}} \\
\cmidrule(lr){3-5} \cmidrule(lr){6-8}
 & & \textbf{S.C.} & \textbf{DQN} & \textbf{\textit{p-value}} & 
     \textbf{S.C.} & \textbf{DQN} & \textbf{\textit{p-value}} \\
\midrule
Regulation & Ratio & $43.22$ & $53.11$ & $0.84$ & $0.094$ & $0.076$ & $0.37$ \\
           & OD    & $4.00$  & $16.02$ & $0.16$ & $0.053$ & $0.072$ & $0.31$ \\
\midrule
Ratio's Tracking & Ratio & $10.10$ & $31.25$ & $0.30$ & $0.11$ & $0.14$ & $0.20$ \\
                 & OD    & --      & --      & --     & $0.05$ & $0.06$ & $0.94$ \\
\midrule
OD's Tracking & Ratio & --      & --      & --     & $0.45$ & $0.13$ & $0.43$ \\
              & OD    & $11.09$ & $10.07$ & $0.66$ & $0.18$ & $0.037$ & $0.36$ \\
\bottomrule
\end{tabular}
\end{table*}

\begin{table*}[ht]
\centering
\caption{Robustness to dilution: comparison between Switching (S.C.) and DQN controllers before and after dilution. Normalized root mean square error (NRMSE) is reported together with paired two-tailed \textit{t}-test $p$-values.}
\label{table:mixing_robustness}
\begin{tabular}{l l
                c c c
                c c c}
\toprule
\multirow{2}{*}{\textbf{Experiment}} & \multirow{2}{*}{\textbf{Signal}} &
\multicolumn{3}{c}{\textbf{Before dilution (NRMSE)}} &
\multicolumn{3}{c}{\textbf{After dilution (NRMSE)}} \\
\cmidrule(lr){3-5} \cmidrule(lr){6-8}
 & & \textbf{S.C.} & \textbf{DQN} & \textbf{\textit{p-value}} & 
     \textbf{S.C.} & \textbf{DQN} & \textbf{\textit{p-value}} \\
\midrule
Robustness & Ratio & $0.066$ & $0.18$ & $0.35$ & $0.067$ & $0.10$ & $0.22$ \\
           & OD    & $0.069$ & $0.057$ & $0.11$ & $0.084$ & $0.077$ & $0.34$ \\
\bottomrule
\end{tabular}
\end{table*}

\section{Discussion and Conclusions}
In this work, building on the dual-chamber control architecture introduced in our previous work \cite{brancato2024ratiometric}, we presented a reliable and robust control architecture for real-time regulation of composition and density of a two-strain bacterial consortium. The control architecture integrates modular control algorithms, allowing controllers to be easily swapped according to experimental and design requirements.
We experimentally implemented and validated the system using two Chi.Bio bioreactors. 
%
We show that the control architecture can  successfully regulate \textit{in vivo} consortium density and composition, requiring only minimal experimental data for model calibration. Specifically, we showed the architecture's ability to steer the consortium to desired setpoints, to track dynamically changing references, and to recover from external perturbations. 
Importantly, the control architecture operated successfully at constant temperature and without genetic modifications, distinguishing this work from previous approaches that require either environmental manipulation \cite{lee2025directing} or engineered auxotrophies \cite{kusuda2021reactor}. 

Several limitations emerged during implementation. The sensing capabilities of the Chi.Bio bioreactors provide only aggregate optical density measurements without real-time, species-specific information. This constraint required an Extended Kalman Filter to reconstruct individual population concentrations, introducing estimation error that contributed to larger steady-state errors in ratio regulation compared to total biomass regulation. Additionally, the peristaltic pumps exhibited backflow effects that limit administrable flow rates and reduce control input resolution at low actuation levels, constraining fine-grained control in scenarios requiring subtle adjustments. These hardware limitations affected all controllers similarly, suggesting that improved sensors and actuators would enhance overall control architecture performance.
Despite these limitations, our experimental validation demonstrates the practical viability of the control architecture. Using all developed controllers we were able to  maintain stable coexistence across all tested scenarios with settling times under one hour and steady-state errors typically under 10\% for total biomass. 

From a control-design perspective, the choice between the model-based and the sim-to-real learning-based controller in the mixing chamber depends primarily on the degree of model uncertainty and the desired control flexibility. The switching controller is preferable when a reasonably accurate process model is available and the actuation space is well defined, as it provides interpretability and guaranteed stability within its design domain. Conversely, the sim-to-real strategy becomes advantageous when the process dynamics are partially known, or when a more complex control law is needed. In the present implementation, the potential advantages of the learning-based controller are partly limited by the use of the model-based observer structure and by the constraints of the Chi.Bio actuation system, which restricts the action spaces. 

Several promising directions exist for future work. Incorporating in situ fluorescence measurements as in \cite{kusuda2021reactor} would eliminate the need for observers, enabling fully state-aware control with direct species-specific feedback. The observer itself could be improved using more robust estimation techniques such as particle filters or moving horizon estimators \cite{pauk2024all,schiller2024moving}, which may better handle nonlinearities and non-Gaussian noise. From a biological perspective, extending the control architecture to control consortia comprising more than two strains would open new avenues for efficient and sustainable production of complex chemical compounds through division-of-labor strategies. Such extensions could leverage the modular control architecture by adding chambers or developing hierarchical control schemes. Furthermore, integrating adaptive or online learning mechanisms could enable long-term autonomous operation, allowing the system to compensate for gradual changes in strain behavior, environmental drift, or culture aging without manual retuning. The sim-to-real pipeline demonstrated here provides a foundation for such adaptive strategies, as mathematical models can be periodically updated using accumulated experimental data.

This work demonstrates that sophisticated feedback control can maintain stable microbial consortia without genetic modifications. The  design of the control architecture allows its use both for fundamental studies and for preliminary steps toward bioprocess applications where genetic interventions may be undesirable due to regulatory constraints or metabolic burden concerns. By providing a versatile control architecture that accommodates both model-based design and data-driven learning, we enable researchers to select control strategies matched to their specific systems, available data, and computational resources, advancing toward more sustainable and efficient bioproduction processes.

\section*{Acknowledgments}
The authors acknowledge support from the lab of prof Diego di Bernardo at the Telethon Institute of Genetics and Medicine (TIGEM) in Naples, Italy, for hosting the experimental set-up that was used for the experiments in this paper and for their invaluable support and collaboration throughout this research project, with special thanks to Barbara Tumaini for her exceptional technical and organizational support. 
\bibliographystyle{IEEEtran}
\bibliography{references}

@article{szotkowski2021bioreactor,
  title={Bioreactor co-cultivation of high lipid and carotenoid producing yeast Rhodotorula kratochvilovae and several microalgae under stress},
  author={Szotkowski, Martin and Holub, Ji{\v{r}}{\'\i} and {\v{S}}imansk{\`y}, Samuel and Huba{\v{c}}ov{\'a}, Kl{\'a}ra and Sikorov{\'a}, Pavl{\'\i}na and Marini{\v{c}}ov{\'a}, Veronika and N{\v{e}}mcov{\'a}, Andrea and M{\'a}rov{\'a}, Ivana},
  journal={Microorganisms},
  volume={9},
  number={6},
  pages={1160},
  year={2021},
  publisher={MDPI}
}

@article{mujtaba2017treatment,
  title={Treatment of real wastewater using co-culture of immobilized Chlorella vulgaris and suspended activated sludge},
  author={Mujtaba, Ghulam and Lee, Kisay},
  journal={Water research},
  volume={120},
  pages={174--184},
  year={2017},
  publisher={Elsevier}
}

@article{bao2023engineering,
  title={Engineering microbial division of labor for plastic upcycling},
  author={Bao, Teng and Qian, Yuanchao and Xin, Yongping and Collins, James J and Lu, Ting},
  journal={Nature Communications},
  volume={14},
  number={1},
  pages={5712},
  year={2023},
  publisher={Nature Publishing Group UK London}
}

@article{kent2002microbial,
  title={Microbial communities and their interactions in soil and rhizosphere ecosystems},
  author={Kent, Angela D and Triplett, Eric W},
  journal={Annual Reviews in Microbiology},
  volume={56},
  number={1},
  pages={211--236},
  year={2002},
  publisher={Annual Reviews 4139 El Camino Way, PO Box 10139, Palo Alto, CA 94303-0139, USA}
}

@article{paerl1996mini,
  title={A mini-review of microbial consortia: their roles in aquatic production and biogeochemical cycling},
  author={Paerl, Hans W and Pinckney, JL},
  journal={Microbial ecology},
  volume={31},
  number={3},
  pages={225--247},
  year={1996},
  publisher={Springer}
}

@article{monod1949growth,
  title={The growth of bacterial cultures},
  author={Monod, Jacques},
  journal={{Annual Review of Microbiology}},
  volume={3},
  number={1},
  pages={371--394},
  year={1949}
}

@article{lee2022cybergenetic,
  title={Cybergenetic control of microbial community composition},
  author={Lee, Ting An and Steel, Harrison},
  journal={Frontiers in Bioengineering and Biotechnology},
  volume={10},
  pages={1873},
  year={2022},
  publisher={Frontiers}
}

@article{chang2023emergent,
  title={Emergent coexistence in multispecies microbial communities},
  author={Chang, Chang-Yu and Baji{\'c}, Djordje and Vila, Jean CC and Estrela, Sylvie and Sanchez, Alvaro},
  journal={Science},
  volume={381},
  number={6655},
  pages={343--348},
  year={2023},
  publisher={American Association for the Advancement of Science}
}

@article{fiore2021feedback,
  title={Feedback ratiometric control of two microbial populations in a single chemostat},
  author={Fiore, Davide and Della Rossa, Fabio and Guarino, Agostino and di Bernardo, Mario},
  journal={IEEE Control Systems Letters},
  volume={6},
  pages={800--805},
  year={2021},
  publisher={IEEE}
}

@article{gutierrez2022dynamic,
  title={Dynamic cybergenetic control of bacterial co-culture composition via optogenetic feedback},
  author={Guti{\'e}rrez Mena, Joaqu{\'\i}n and Kumar, Sant and Khammash, Mustafa},
  journal={Nature Communications},
  volume={13},
  number={1},
  pages={4808},
  year={2022},
  publisher={Nature Publishing Group UK London}
}

@article{bertaux2022enhancing,
  title={Enhancing bioreactor arrays for automated measurements and reactive control with ReacSight},
  author={Bertaux, Fran{\c{c}}ois and Sosa-Carrillo, Sebasti{\'a}n and Gross, Viktoriia and Fraisse, Achille and Aditya, Chetan and Furstenheim, Mariela and Batt, Gregory},
  journal={Nature communications},
  volume={13},
  number={1},
  pages={3363},
  year={2022},
  publisher={Nature Publishing Group UK London}
}

@article{kusuda2021reactor,
  title={Reactor control system in bacterial co-culture based on fluorescent proteins using an Arduino-based home-made device},
  author={Kusuda, Minori and Shimizu, Hiroshi and Toya, Yoshihiro},
  journal={Biotechnology Journal},
  volume={16},
  number={12},
  pages={2100169},
  year={2021},
  publisher={Wiley Online Library}
}

@article{steel2019chi,
  title={Chi. Bio: An open-source automated experimental platform for biological science research},
  author={Steel, Harrison and Habgood, Robert and Kelly, Ciar{\'a}n and Papachristodoulou, Antonis},
  journal={BioRxiv},
  pages={796516},
  year={2019},
  publisher={Cold Spring Harbor Laboratory}
}

@inproceedings{paradowski2017approach,
  title={An approach to determine observability of nonlinear systems using interval analysis},
  author={Paradowski, Thomas and Tibken, Bernd and Swiatlak, Robert},
  booktitle={2017 American Control Conference (ACC)},
  pages={3932--3937},
  year={2017},
  organization={IEEE}
}

@article{lugagne2017balancing,
  title={Balancing a genetic toggle switch by real-time feedback control and periodic forcing},
  author={Lugagne, Jean-Baptiste and Sosa Carrillo, Sebasti{\'a}n and Kirch, Melanie and K{\"o}hler, Agnes and Batt, Gregory and Hersen, Pascal},
  journal={Nature communications},
  volume={8},
  number={1},
  pages={1--8},
  year={2017},
  publisher={Nature Publishing Group}
}

@article{steel2020situ,
  title={In situ characterisation and manipulation of biological systems with Chi. Bio},
  author={Steel, Harrison and Habgood, Robert and Kelly, Ciar{\'a}n L and Papachristodoulou, Antonis},
  journal={PLOS Biology},
  volume={18},
  number={7},
  pages={e3000794},
  year={2020},
  publisher={Public Library of Science San Francisco, CA USA}
}

@inproceedings{tani2019,
  title={A hybrid control against species invasion in the chemostat},
  author={Tani, Fatima-Zahra and others},
  booktitle={Proc. of the IEEE Conference on Decision and Control},
  pages={2814--2819},
  year={2019}
}

@book{bastin2013line,
  title={On-line estimation and adaptive control of bioreactors},
  author={Bastin, Georges and Dochain, D},
  year={1990},
  address={Amsterdam, Netherlands},
  publisher={Elsevier Science Publisher}
}

@article{mnih2015human,
  title={Human-level control through deep reinforcement learning},
  author={Mnih, Volodymyr and Kavukcuoglu, Koray and Silver, David and Rusu, Andrei A and Veness, Joel and Bellemare, Marc G and Graves, Alex and Riedmiller, Martin and Fidjeland, Andreas K and Ostrovski, Georg and others},
  journal={nature},
  volume={518},
  number={7540},
  pages={529--533},
  year={2015},
  publisher={Nature Publishing Group}
}

@article{litcofsky2012iterative,
  title={Iterative plug-and-play methodology for constructing and modifying synthetic gene networks},
  author={Litcofsky, Kevin D and Afeyan, Raffi B and Krom, Russell J and Khalil, Ahmad S and Collins, James J},
  journal={Nature methods},
  volume={9},
  number={11},
  pages={1077--1080},
  year={2012},
  publisher={Nature Publishing Group US New York}
}

@article{diender2021synthetic,
  title={Synthetic co-cultures: novel avenues for bio-based processes},
  author={Diender, Martijn and Olm, Ivette Parera and Sousa, Diana Z},
  journal={Current Opinion in Biotechnology},
  volume={67},
  pages={72--79},
  year={2021},
  publisher={Elsevier}
}

@article{martinez2023optimal,
  title={Optimal protein production by a synthetic microbial consortium: Coexistence, distribution of labor, and syntrophy},
  author={Mart{\'\i}nez, Carlos and Cinquemani, Eugenio and Jong, Hidde de and Gouz{\'e}, Jean-Luc},
  journal={Journal of Mathematical Biology},
  volume={87},
  number={1},
  pages={23},
  year={2023},
  publisher={Springer}
}

@inproceedings{asswad2024optimization,
  title={Optimization of microalgae biosynthesis via controlled algal-bacterial symbiosis},
  author={Asswad, Rand and Djema, Walid and Bernard, Olivier and Gouz{\'e}, J-L and Cinquemani, Eugenio},
  booktitle={2024 IEEE 63rd Conference on Decision and Control (CDC)},
  pages={589--594},
  year={2024},
  organization={IEEE}
}

@inproceedings{brancato2024vivo,
  title={In vivo learning-based control of microbial populations density in bioreactors},
  author={Brancato, Sara Maria and Salzano, Davide and De Lellis, Francesco and Fiore, Davide and Russo, Giovanni and di Bernardo, Mario},
  booktitle={6th Annual Learning for Dynamics \& Control Conference},
  pages={941--953},
  year={2024},
  organization={PMLR}
}

@article{zhou2015distributing,
  title={Distributing a metabolic pathway among a microbial consortium enhances production of natural products},
  author={Zhou, Kang and Qiao, Kangjian and Edgar, Steven and Stephanopoulos, Gregory},
  journal={Nature biotechnology},
  volume={33},
  number={4},
  pages={377--383},
  year={2015},
  publisher={Nature Publishing Group US New York}
}

@article{brancato2024ratiometric,
  title={Ratiometric control of two microbial populations via a dual chamber bioreactor},
  author={Brancato, Sara Maria and Salzano, Davide and Fiore, Davide and Russo, Giovanni and di Bernardo, Mario},
  journal={IEEE Control Systems Letters},
  year={2024},
  publisher={IEEE}
}

@article{pauk2024all,
  title={An all-in-one state-observer for protein refolding reactions using particle filters and delayed measurements},
  author={Pauk, Jan Niklas and Igwe, Chika Linda and Herwig, Christoph and Kager, Julian},
  journal={Chemical Engineering Science},
  volume={287},
  pages={119774},
  year={2024},
  publisher={Elsevier}
}

@article{schiller2024moving,
  title={Moving horizon estimation for nonlinear systems with time-varying parameters},
  author={Schiller, Julian D and M{\"u}ller, Matthias A},
  journal={IFAC-PapersOnLine},
  volume={58},
  number={18},
  pages={341--348},
  year={2024},
  publisher={Elsevier}
}

@article{lee2025directing,
  title={Directing microbial co-culture composition using cybernetic control},
  author={Lee, Ting An and Morlock, Jan and Allan, John and Steel, Harrison},
  journal={Cell Reports Methods},
  volume={5},
  number={3},
  year={2025},
  publisher={Elsevier}
}

@INPROCEEDINGS{7983780,
  author={Spielberg, S.P.K. and Gopaluni, R.B. and Loewen, P.D.},
  booktitle={2017 6th International Symposium on Advanced Control of Industrial Processes (AdCONIP)}, 
  title={Deep reinforcement learning approaches for process control}, 
  year={2017},
  volume={},
  number={},
  pages={201-206},
  doi={10.1109/ADCONIP.2017.7983780}}

\section*{Appendix A: Observer Design Details}

\subsection*{Observability Analysis}
Consider the dynamical system
\begin{equation} \label{eq:nonlinear_sys}
    \begin{split}
        \dot x =& f(x,u) \\
        y =& h(x)
    \end{split}
\end{equation}
where $x\in \mathbb{R}^n$ is the state of the system, $u\in \mathbb{R}^m$ are the inputs, $y \in \mathbb{R}^p$ are the output of the system, $f:\mathbb{R}^n \times \mathbb{R}^m \to \mathbb{R}^n$ is the vector field describing the system dynamics, and $h: \mathbb{R}^n \to \mathbb{R}^p$ is the function describing the available measures.
To characterize the observability of system \eqref{eq:nonlinear_sys}, we constructed the observability matrix. We defined the observability function
\begin{equation}
q(x(t)) = \begin{pmatrix} h(x(t)) \\ \mathcal{L}_f h \end{pmatrix},
\end{equation}
where $h(x)$ is a function describing the available measures and $\mathcal{L}_{f}h$ is the Lie derivative of $h(x)$ along the system dynamics. Then we computed $\mathcal{O} = \frac{\partial q(x(t))}{\partial x}$ \cite{paradowski2017approach}, which, for model \eqref{eq:simplified-model}, is
\begin{equation}
\mathcal{O} = \begin{pmatrix}
1 & 1 \\
(\mu^*_1 - (D_1 + D_2)) & (\mu^*_2 - (D_1 + D_2))
\end{pmatrix}.
\end{equation}
Since $\mu^*_1 \neq \mu^*_2$ for non-complementary populations, the system is locally observable.

\subsection*{EKF Parameter Tuning}
The noise covariances ($\mathbf{Q}$, $\mathbf{R}$) and initial uncertainty ($\mathbf{P}_0$) were tuned using a genetic algorithm to minimize squared error between EKF estimates and experimental data. Optimal values were $\mathbf{Q} = 10^{-5}\mathbf{I}$, $\mathbf{R} = 5$, and $\mathbf{P}_0 = 0.072\mathbf{I}$, where $\mathbf{I}$ denotes the identity matrix of appropriate dimension.

\subsection*{Flow Cytometry Protocol}
Samples from the mixing chamber were analyzed using flow cytometry. After excluding debris and cell agglomerates using FSC and SSC channels, cells were classified by fluorescence. Cells with FL1-H $\geq 2 \times 10^3$ (on log scale) were classified as strain 2 (expressing GFP). Remaining cells were classified as strain 1.

\end{document}